\documentclass[conference]{IEEEtran}
\IEEEoverridecommandlockouts
\usepackage{cite}
\usepackage{amsmath,amssymb,amsfonts}
\usepackage{algorithmic}
\usepackage{graphicx}
\usepackage{textcomp}
\usepackage{xcolor}
\usepackage{dblfloatfix}

\def\BibTeX{{\rm B\kern-.05em{\sc i\kern-.025em b}\kern-.08em
    T\kern-.1667em\lower.7ex\hbox{E}\kern-.125emX}}
\begin{document}

\title{IoT Cloud RAN Testbed for Ultra-Precise TDoA-based Localization in LPWANs}

\author{
\IEEEauthorblockN{Thomas Maul, Joerg Robert}
\IEEEauthorblockA{\textit{Technische Universität Ilmenau} \\
\textit{M2M Research Group}\\
Ilmenau, Germany \\
\{thomas.maul, joerg.robert\}@tu-ilmenau.de}
\and
\IEEEauthorblockN{Sebastian Klob}
\IEEEauthorblockA{\textit{Friedrich-Alexander Universität Erlangen-Nürnberg (FAU)} \\
\textit{Information Technology (Communication Electronics)}\\
Erlangen, Germany \\
sebastian.klob@fau.de}}

\IEEEaftertitletext{This work has been submitted to the IEEE for possible publication. Copyright may be transferred without notice, after which this version may no longer be accessible.}
\maketitle

\begin{abstract}
There have been many research efforts in the area of localization in recent years. 
Especially within the Internet of Things (IoT), the knowledge of position information for individual components is of great interest, for example, in asset tracking, to name just one. 
However, many of these use cases require a high energy efficiency, making a GNSS-based approach infeasible. 
One promising candidate can be found in Low Power Wide Area Networks (LPWAN), which enable battery lifetimes of up to 20 years. However, no gold standard for localization exists for these types of networks. Our work proposes a testbed architecture that allows the investigation and development of localization algorithms within LPWA Networks. 
The concept is built on a Cloud Radio Access Network (CRAN) architecture that allows the streaming of IQ from remote base stations to a central processing unit. Furthermore, the architecture is expanded by a synchronization concept based on Signals of Opportunity (SoO) to enable the testbed for runtime-based positioning. 
\\Therefore, we propose a hardware concept consisting of antennas and a low-cost off-the-shelf software-defined radio (SDR)-based frontend architecture and a software framework using a hypertext transfer protocol (HTTP)-based server and client architecture. 
The proposed system is installed in an urban environment. Initial measurements are conducted, where it can be shown that the proposed architecture can be used for highly precise Time Difference of Arrival (TDoA) measurements, offering the possibility of time synchronization down to approximately 200\, ps and frequency synchronization of 3\,mHz.
\\ 
\end{abstract}

\section{Introduction}
In recent years, the demand for IoT devices and solutions has increased sharply \cite{IoT_growth}. One use case that gained increasing attention is asset tracking in modern industrial processes, where the knowledge of position information helps to optimize manufacturing processes.
There exists a variety of different localization solutions \cite{IoT_Loc}.
One promising candidate can be found in Low Power Wide Area Networks (LPWANs) since classical localization strategies like Global Navigation Satellite Systems (GNSS) are often not feasible due to cost and energy efficiency limitations. 
The most prominent representatives of this category of communication systems are LoRa\footnote{https://www.semtech.com/lora (accessed December 2023)}, sigfox\footnote{https://www.sigfox.com/ (accessed December 2023)}, and ETSI TS 103 357 \cite[Sec.~6]{ETSI} alias mioty\footnote{https://mioty-alliance.com/ (accessed December 2023)}.
For example, \cite{Sigfox}, and \cite{NBIoT} present techniques based on the mentioned LPWAN solutions that solely rely on the measurement of Received Signal Strength Indicator (RSSI) values, allowing localization with little complexity.
However, this approach only delivers an accuracy of hundreds of meters in larger networks. 
A more compromising method is the exploitation of Time Difference of Arrival (TDoA) values for localization \cite{TDoA_0}. This approach generally delivers much more accurate position estimates than RSSI-based solutions. 
However, a major drawback is the necessity of ultra-precise synchronization between the base stations.
The proposal in \cite{TDoA_1} therefore uses a GNSS-based approach, which results in larger synchronization errors, negatively affecting the localization accuracy.

This paper proposes a testbed based on a Cloud Radio Access Network (CRAN) architecture suitable for the LoRa, sigfox, and mioty signals, building upon the work in~\cite{Testbed_Indoor}.
In addition, it enables TDoA-based localization by adding the possibility of synchronization of the remote base stations with the use of broadcast signals as Signals of Opportunity (SoO), as described in \cite{Synch_Basics} and \cite{Synch}. The stated testbed architecture can digitize the received antenna signals and stream them from a remote base station to a centralized processing unit, where they are fully synchronized in frequency and time and can be further processed in TDoA localization algorithms. 

The remainder of this document is structured as follows:
 Section \ref{Architecture} gives an overview of the architecture and each component. Section \ref{Ilmenau} shows the installation within a real-world environment in an urban area. Section \ref{Measurements} gives insight into initial measurement results and Section \ref{Conclusion} concludes this document.

\section{Proposed Testbed Architecture} \label{Architecture}
\begin{figure}[htbp]
\centerline{\includegraphics[width=0.47\textwidth]{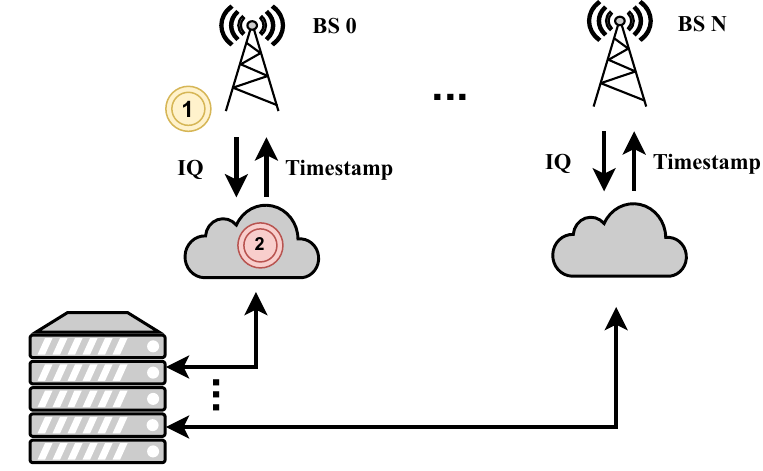}}
\caption{Cloud RAN architecture consisting of a main server and several remote base stations. The main server requests data at a certain timestamp and the base station answers by transmitting the IQ data.}
\label{CRAN_Overview}
\end{figure}

This chapter introduces the CRAN architecture as it is implemented within this testbed. Therefore, we start by considering a classical CRAN approach, visualized in Fig. \ref{CRAN_Overview}.
The architecture is based on a main computer that fulfills several functions. Both the synchronization and localization algorithms are deployed here. Further, the main server coordinates the request and storage of IQ data. Therefore, the server sends a request to a remote base station containing the exact timestamp the data is required. The base station answers this request by transmitting the IQ data to the main server. However, to implement this architecture, two points are of crucial importance. At first, a suitable base station hardware is required. This includes antennas, the frontend, and a radio link to establish a stable connection between the remote station and the main server. Secondly, a framework is required that manages the requests and initiates the transmission of the digitized signals from the antennas. In the following, we will closely examine both critical components of the proposed CRAN architecture.

\subsection{Base Station Hardware}
\begin{figure}[htbp]
\centerline{\includegraphics[width=0.47\textwidth]{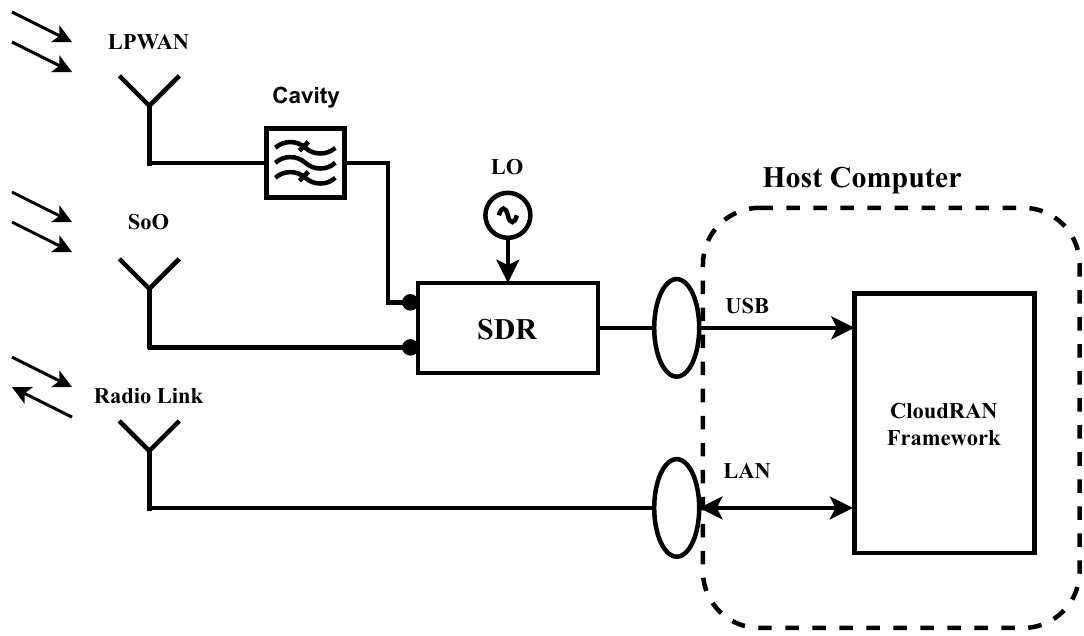}}
\caption{Proposed Hardware Concept for the remote base stations containing the antennas, bandpass filter, and the SDR-based frontend for digitization.}
\label{Hardware}
\end{figure}

In the context of a CRAN architecture, the task of the base station is receiving signals in the desired frequency range in combination with digitization and pre-processing of the IQ data before transmission to the central server. Fig. \ref{Hardware} shows the proposed hardware setup at the base station.
\\In the case at hand, two channels have to be received, namely the LPWAN data for localization and the SoO data for synchronization. Both signals are fed into the receiving SDR-based front end. However, since most LPWANs operate in the license-exempt frequency band at 868\,MHz, many communication systems exist in neighboring channels, potentially causing severe interference. Strong signals from cellular broadcast stations like GSM or LTE can lead to a reduced dynamic range in the receiving front end. To prevent this, we add a bandpass filter with a narrow passband and a high attenuation in the stopband. The bandpass filter, designed as a so-called cavity filter\footnote{https://amphenolprocom.com/de/produkte/filters-de/930-bpf-900-3 (accessed December 2023)}, suppresses these potentially harmful signals.
In the prevailing case, the cavity filter is tuned to a passband of 863\,MHz - 870\,MHz. Another important component is the SDR-based frontend. In the case at hand, we used the SDRPlay RSPDuo\footnote{https://www.sdrplay.com/rspduo/ (accessed December 2023)} device, featuring a low noise figure of $F_{\text{RX}}$ = 2\,dB and low cost (250\$). However, this device can be replaced by any off-the-shelf front end. The only requirement is the LO-sharing capability, where one oscillator feeds both receiving channels. This requirement is a direct consequence of the synchronization approach \cite{Synch}. 
In addition, the RSPDuo is clocked by an external oscillator\footnote{https://www.digikey.de/de/products/detail/connor-winfield/OH200-61003CF-024-0M/15291389 (accessed December 2023)}. In this case, it is a very frequency-stable oven-controlled crystal oscillator with an excellent phase-noise profile to improve synchronization and localization accuracy.
A USB interface connects the front end to the host computer, whose task is to process the digitized antenna signals and handle the IQ requests from the central server. This is also where the CRAN framework is running. The radio link is the last crucial component of the hardware setup, which guarantees a stable and fast connection between the central server and the remote base station. The link\footnote{https://eu.store.ui.com/eu/en/products/airfiber-60-lr (accessed December 2023)} consists of two antennas, which are set up at the locations to be connected and aligned with each other. The used link is bidirectional and capable of bridging distances up to 12\,km with a capacity up to 2\,GBit/s.

\subsection{Cloud RAN Framework}
\label{CloudRAN}
\begin{figure}[htbp]
\centerline{\includegraphics[width=0.47\textwidth]{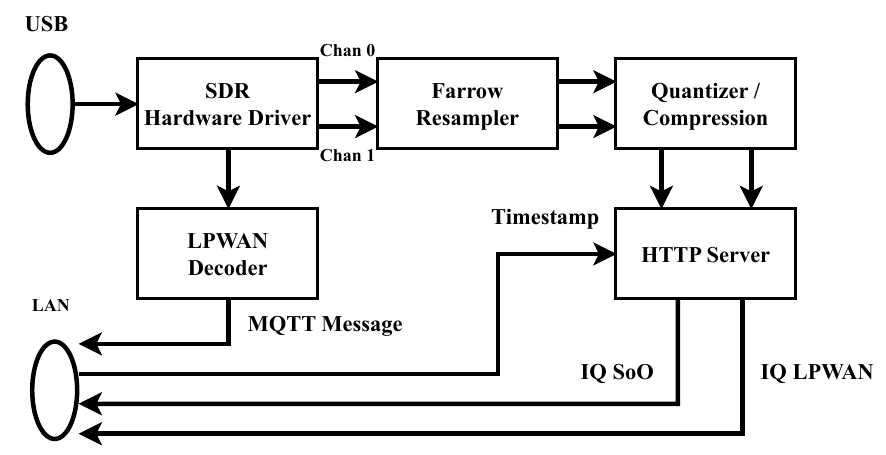}}
\caption{Proposed Cloud RAN Software Framework with individual components for handling the IQ data requests.}
\label{Framework}
\end{figure}

The implementation of our proposed CRAN framework is based on the work presented in \cite{DFCPP}. Fig. \ref{Framework} shows an overview of the CRAN concept with individual components. 

The main idea behind the framework is to do all the signal processing in blocks. This allows individual blocks to be easily exchanged or removed without the need for complex adjustments within the framework.
\\The first block represents the hardware driver for the SDR. In the case of the RSPDuo, an implementation of SoapySDR\footnote{https://github.com/pothosware/SoapySDR (accessed December 2023)} is used. The driver directly passes important settings like gain, center frequency, bandwidth, and sampling rate to the SDR. The used sampling rate here is $f_s$ = 2.0\,MHz with a bandwidth of B = 1.536\,MHz for both channels. The driver also distributes the data to downstreamed blocks.
One path directly feeds the LPWAN decoder, which in our case complies with the mioty standard defined in ETSI TS 103 357 \cite[Sec. 6]{ETSI}.
The decoder estimates multiple parameters like the message payload, reception time, and RSSI value. The decoder generates a "Message Queuing Telemetry Transport" (MQTT) message and propagates it to the main station, where the information can be used for further processing in the localization algorithm. The MQTT message format is specially for IoT use cases designed format, optimized for short messages.

In the second path, the digitized IQ data from both channels is fed into a fractional resampling block, designed as a Farrow structure~\cite{Farrow}.
It resamples both channels to a sampling rate of $f_{\text{res}} = 2^{21}\,\text{Hz}$.
This number of two values allows a simplified Fast Fourier Transform (FFT) calculation, which is important in the following block. Here, the spectrum of the data is segmented according to the frequency segmentation principles in \cite{RAN}.
Spectrum segmentation and applying compression like Huffman coding or Lempel-Ziv encoding drastically reduces the amount of data that needs to be transmitted over the radio link. 
A strong reduction can be achieved, especially with sparse spectra, as is common with LPWANs. 
Nevertheless, the quality of synchronization and localization could decline when applying strict segmentation or compression. 
This is why we additionally implemented direct quantization with an adjustable number of bits. This is also the preferred way for broadband signals that cannot be compressed significantly.
Possible resolutions are 8-bit or 16-bit, resulting in a trade-off between the volume of data and the achieved dynamic range. 

The last block in the chain of the CRAN framework is the HTTP server.
This is where the requests from the main server arrive, encoded as HTTP requests. The message contains the timestamp for the desired IQ data and the frontend channel. When using spectrum segmentation, the message additionally contains the subband. The framework then processes the request and searches for the corresponding block of IQ data and, when a subband was specified, the correct part of the spectrum. The server then replies by transmitting the requested IQ data or an error code should the data be unavailable. If any compression or segmentation is applied, this must be undone at this point to recover the original data. The client repeats this procedure until the required period of data has been successfully requested.

 \section{Testbed Installation in Ilmenau} \label{Ilmenau}
\begin{figure}[htbp]
\centerline{\includegraphics[width=0.47\textwidth]{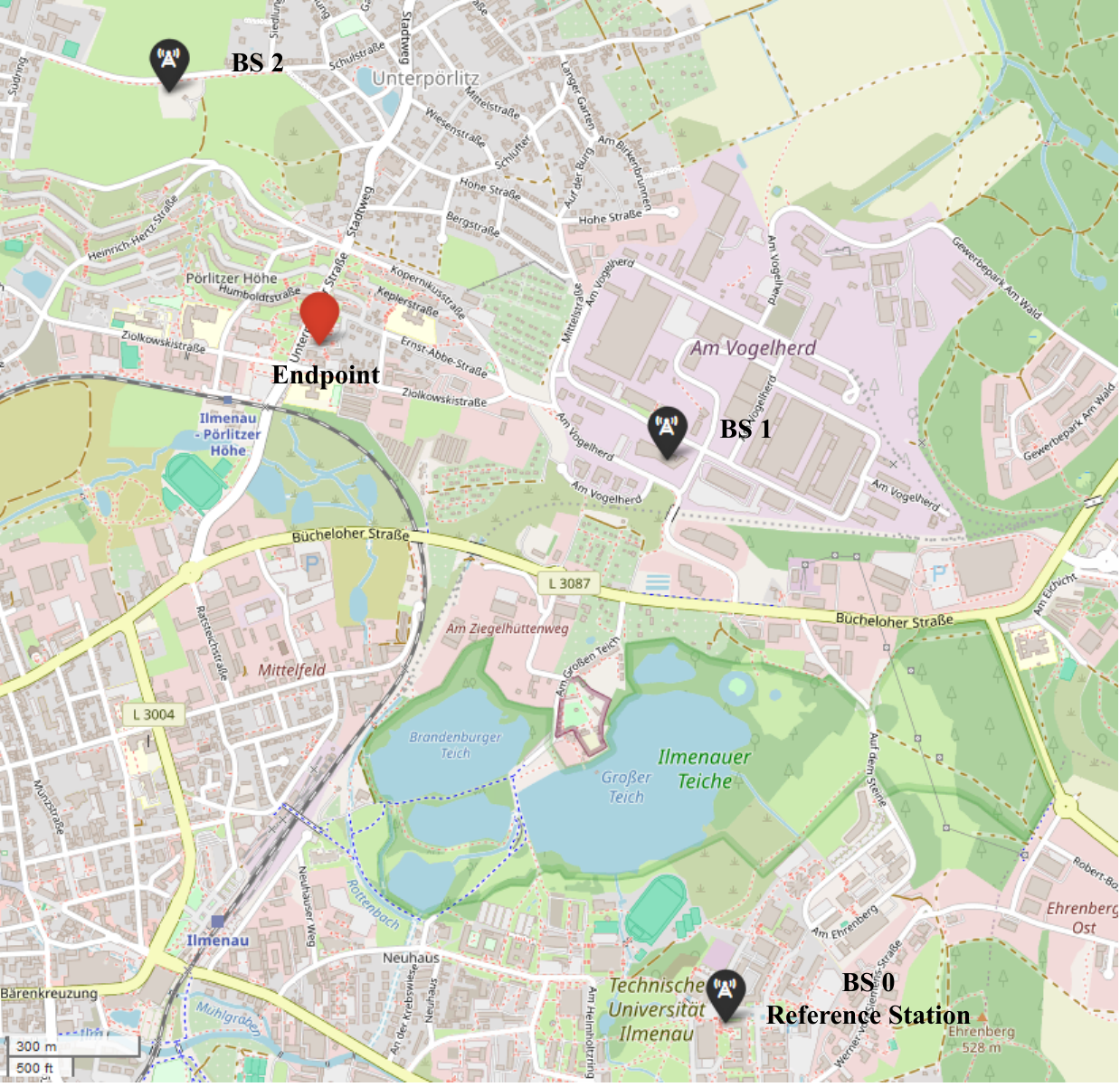}}
\caption{Map of the  testbed installation in Ilmenau. Three base stations (black icons) cover the whole city area, whereas BS 0 denotes the reference station. One endpoint for testing purposes is marked with a red icon. ©~OpenStreetMap-Contributers }
\label{MAP}
\end{figure}

This section shows the installation of the proposed CRAN Testbed in a real-world environment. The installation is located in Ilmenau, a city in Thuringia, Germany. In the initial expansion stage, the testbed consists of three base stations distributed throughout the city. Fig. \ref{MAP} shows the scale and the position of the remote base stations within the city. 

\begin{figure}[htbp]
\centerline{\includegraphics[width=0.47\textwidth]{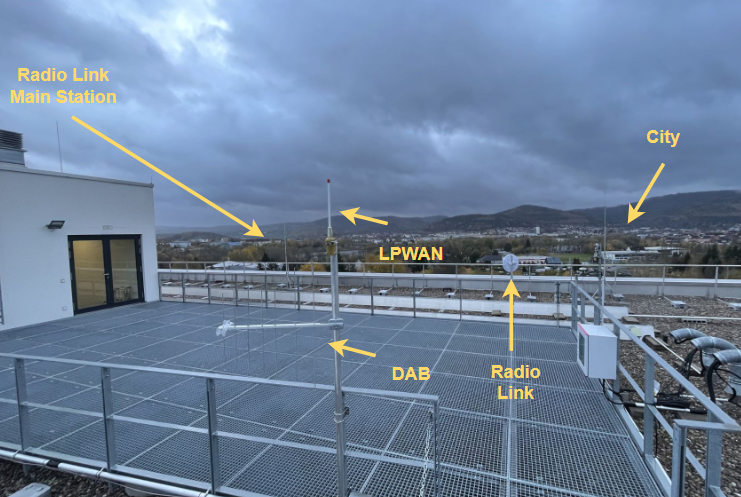}}
\caption{BS 1 during the installation process on the roof of a building in Ilmenau. The picture includes the antennas for LPWAN as well as for DAB. In the background, the radio link is visible, which is precisely aligned to the counterpart at the reference station (BS 0)}
\label{Installation}
\end{figure}

The distance between the reference station and the first and second base station is about 1.3\,km and 2.4\,km, respectively. In order to establish a connection between the reference station, where the main server is located, and the remote base stations, we set up two independent links from the reference station to the respective remote base station. Fig. \ref{Installation} shows the setup of the remote base station BS 1 during the installation process with all relevant components.


The installation of BS 1 is located on the roof of a three-story building on the edge of the city center of Ilmenau. The setup consists of the LPWAN antenna\footnote{https://shop.pro-tecs.de/main-categories/antennen/feststations-maritim-antennen/cxl-900-1lw-l.html (accessed December 2023)} for communication and localization and the DAB antenna\footnote{https://stecker-shop.net/LP7902G7 (accessed December 2023)} for synchronization. The radio link can be seen at the edge of the railing in the background. It is precisely aligned with the remote station at the reference station. The frontend and the computer are located in the server room below the roof.

\section{Initial Measurement Results}\label{Measurements}
This chapter shows initial measurement results to evaluate the overall concept for functionality. This analysis concerns the data transfer of the LPWAN data, including the spectrum segmentation and quantization on the server side, as well as the functionality of the LPWAN decoder. In a further section, we will investigate the performance of the time and frequency synchronization using Signals of Opportunity.
\subsection{LPWAN}
\begin{figure}[htbp]
\centerline{\includegraphics[width=0.47\textwidth]{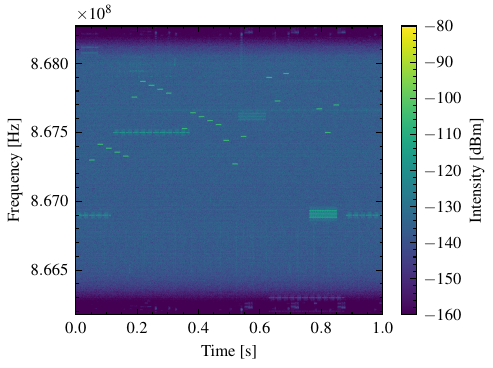}}
\caption{Spectrogram of the IQ data received at BS 0, the data are quantized with a resolution of 16\,Bit.}
\label{LPWAN_SPEC}
\end{figure}

In the first step, we will examine the spectrum segmentation of the proposed CRAN framework. To test the quantization only, we will refrain from compressing the data and quantizing the entire spectrum with a resolution of 16\,Bits without any subband segmentation. Fig. \ref{LPWAN_SPEC} shows the requested IQ data of the endpoint from Fig. \ref{MAP} for the reference station (BS0), plotted as spectogram.
\\Fig. \ref{LPWAN_SPEC} shows the spectrum of a mioty telegram with the easily recognizable Telegram Splitting Multiple Access (TSMA) structure. The entire telegram is divided into individual packets randomly distributed in frequency and time to increase robustness against interference \cite{TSMA}. Furthermore, the spectrum does not show any quantization artifacts or data loss. Nevertheless, we can recognize co-channel interference from other communication systems in the same frequency band, just like the filter's edges in the frontend at the edges of the spectrum.
\\Another critical aspect of the proposed system is the functionality of the LPWAN decoder. As Section \ref{CloudRAN} mentions, the decoder directly gets the raw IQ data from the CRAN framework. Therefore, the decoder's performance depends directly on the quality of the IQ data provided by the framework. A good metric for assessing the quality here would, therefore, be a long-term measurement in which many packets are received over a more extended period of time. Table \ref{LPWAN_Table}, thus, shows the packet error rate (PER) and the average RSSI for all three base stations. The node was placed within the testbed according to Fig. \ref{MAP} and transmitted data every 90\,s over a total duration of 9\,h.

\begin{table}[!h]
\caption{\label{LPWAN_Table}Examination of the decoder quality over a period of 9\,h}
\begin{center}
\begin{tabular}{|c|c|c|}
\hline
BS & packets received [\%] & avg. RSSI [dBm]\\
\hline
\hline
0    &  100.0 & -96.65 \\
\hline
1    &  100.0 & -120.56 \\
\hline
2    &  100.0 & -115.544 \\
\hline
\end{tabular}
\end{center}

\end{table}
Table \ref{LPWAN_Table} clearly shows that all three base stations were able to decode all transmitted packets, which proves the correct functioning of the decoder. This is particularly important as the information from the decoder can later be useful for developing localization algorithms. For example, the exact reception time can be used to request the IQ data of the respective base station at the correct timestamp.

\subsection{Synchronization}
\begin{figure}[htbp]
\centerline{\includegraphics[width=0.47\textwidth]{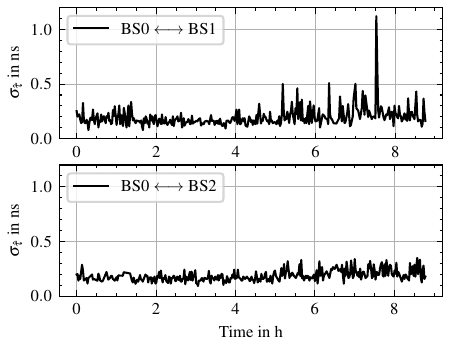}}
\caption{Measured standard deviation $\sigma_{\hat{\tau}}$ as a function of time for the time synchronization between the reference station (BS 0) and both remote stations (BS1, BS2).}
\label{SCO}
\end{figure}

\begin{figure}[htbp]
\centerline{\includegraphics[width=0.47\textwidth]{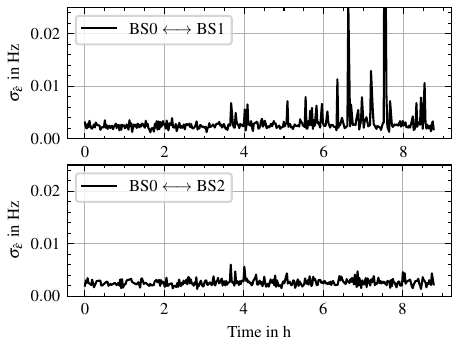}}
\caption{Measured standard deviation $\sigma_{\hat{\epsilon}}$ as a function of time for the frequency synchronization between the reference station (BS 0) and both remote stations (BS1, BS2).}
\label{CFO}
\end{figure}

Another critical investigation concerns synchronization accuracy between the base stations, which states a fundamental requirement for developing precise runtime-based positioning algorithms. In this paper, we refer to the synchronization algorithm examined in more detail in \cite{Synch}. The proposed algorithm uses LO-sharing frontends with two channels in the base stations and, therefore, receives the SoO parallel to the LPWAN waveform. Afterward, the synchronization algorithm is applied to the SoO data streams of neighboring base stations. The synchronization parameters can then be calculated from the cross-correlation function (CCF). The estimated parameters, namely Carrier Frequency Offset (CFO), Sampling Clock Offset (SCO), and time synchronization can then be used to correct the synchronization errors on the LPWAN waveform. However, \cite{Synch} only investigated the achievable performance in a cable-based laboratory setup. With the testbed architecture presented in this document, we now have the opportunity to carry out the synchronization in a real-world environment with spatially separated base stations. We select the algorithm parameters identically to achieve comparability with the results obtained in \cite{Synch}. The only deviation is the higher signal-to-noise ratio ($E_S/N_0 = 24\,\text{dB}$) in the prevailing case due to using a directional DAB antenna. 
\\The starting point for the evaluation is again the endpoint shown in Fig. \ref{MAP}. After each successfully decoded telegram, both the SoO and the LPWAN data are requested from remote base stations with a resolution of 16 Bit. The aggregated data at the main server is then exposed to the synchronization algorithm, and the evaluation is conducted. 
\\ We start the evaluation by inspecting the accuracy of the time synchronization. Fig. \ref{SCO} shows the progression of the estimated standard deviation $\sigma_{\hat{\tau}}$ over time for the synchronization between the base stations with index 0 and 1 and between base stations 0 and 2.
\\ The evaluation of the standard deviations from Fig. \ref{SCO} shows that the accuracy of the time synchronization reaches values down to 1\,ns ($\approx$ 0.3\,m, considering the speed of light). In detail, the average standard deviation is found at approximately 194\,ps (0.194\,m) and 187\,ps (0.188\,m) respectively. The accuracies achieved are very promising, even for ultra-precise runtime-based localization.
\\ An analog evaluation based on the identical data set can then be carried out for frequency synchronization. Therefore, Fig. \ref{CFO} shows the progression of the estimated standard deviation $\sigma_{\hat{\epsilon}}$ of the frequency estimation over time, again evaluated for between the reference station (BS 0) and the corresponding remote base station.

The information from Fig. \ref{CFO} can be summarized similarly like Fig. \ref{SCO}. Very low standard deviations can be achieved for both courses. In detail, the standard deviations are found down to 3.1\,mHz and 2.65\,mHz respectively. However, in contrast to the time synchronization, the results here depend on the underlying carrier frequency. The carrier frequency for the waveform used here is $f_{\text{SoO}}$ = 178.352\,MHz, for higher frequencies, the expected standard deviation increases linearly accordingly. 

\section{Summary and Conclusion}
\label{Conclusion}
This article presented an approach for a Cloud Radio Access Network (CRAN)-based testbed for Time Difference of Arrival (TDoA) localization within Low Power Wide Area Networks (LPWAN). The proposed architecture comprises multiple remote base stations and one central reference station, which can be precisely synchronized in frequency and time using a broadcast (DAB) as Signals of Opportunity (SoO). Each base station comprises a hardware concept including antennas, a radio link for data transfer, and a low-cost off-the-shelf frontend architecture. Furthermore, the presented concept includes a software framework with an LPWAN decoder and the possibility of requesting IQ data from a selected base station and streaming it to the reference station. The proposed testbed architecture was set up in an urban environment. Initial measurement results have proven the functionality of the IQ data exchange and the LPWAN decoder. Furthermore, it has been shown that accuracy in the range of 1\,ns for time synchronization and 3\,mHz for frequency synchronization is possible, enabling the testbed for ultra-precise positioning. However, many other applications are also conceivable here. The flexibly expandable concept could be used for applications in Integrated Sensing and Communication (ICAS) applications, to name just one.
\\ In future work, we will evaluate TDoA-based localization algorithms in the constructed testbed. Other focal points include further investigations into the achievable synchronization accuracy. Furthermore, a reduction of the data rate for transmission of IQ data can be achieved by exploiting compression. The significantly reduced data rate could make the separate radio link obsolet and enable IQ data to be exchanged via the open Internet.
\section*{Acknowledgment}

This work is part of the research project 5G-Flexi-Cell (grant no. 01MC22004B) funded by the German Federal Ministry for Economic Affairs and Climate Action (BMWK) based on a decision taken by the German Bundestag.


\end{document}